\documentclass[doublespacing]{elsart}
\usepackage{epsfig}

\begin{document}

\begin{frontmatter}

\title{Application of PbWO$_4$ crystal scintillators in experiment to
search for $2\beta$ decay of $^{116}$Cd}

\author[INR-Kiev]{F.A.~Danevich\thanksref{1}}
 \thanks[1]{Corresponding author.
 Address: Institute for Nuclear Research, Prospect Nauki 47, MSP 03680 Kiev, Ukraine;
 tel: +380-44-265-1111;
 fax: +380-44-265-4463;
 email address: danevich@kinr.kiev.ua},
\author[INR-Kiev]{A.Sh.~Georgadze},
\author[INR-Kiev]{V.V.~Kobychev},
\author[INR-Kiev]{B.N.~Kropivyansky},
\author[INR-Kiev]{S.S.~Nagorny},
\author[INR-Kiev]{A.S.~Nikolaiko},
\author[INR-Kiev]{D.V.~Poda},
\author[INR-Kiev]{V.I.~Tretyak},
\author[INR-Kiev]{S.S.~Yurchenko}

\address[INR-Kiev]{Institute for Nuclear Research, MSP 03680 Kiev, Ukraine}

\author[ISC]{B.V.~Grinyov},
\author[ISC]{L.L.~Nagornaya},
\author[ISC]{E.N.~Pirogov},
\author[ISC]{V.D.~Ryzhikov}

\address[ISC]{Institute for Scintillation Materials, 61001 Kharkov, Ukraine}

\author[JINR]{V.B.~Brudanin}

\address[JINR]{Joint Institute for Nuclear Research, 141980 Dubna, Russia}

\author[INP-Minsk]{M.~Fedorov},
\author[INP-Minsk]{M.~Korzhik},
\author[INP-Minsk]{A.~Lobko},
\author[INP-Minsk]{O.~Miussevitch}

\address[INP-Minsk]{Research Institute of Nuclear Problems,  220050 Minsk, Belarus}

\author[IM]{I.M.~Solsky}

\address[IM]{Institute for Materials, 79031 Lviv, Ukraine}

\begin{abstract}
PbWO$_4$ crystal scintillators are discussed as an active shield and light-guides in
$^{116}$Cd double beta decay experiment with CdWO$_4$ scintillators.
Scintillation properties and radioactive contamination of PbWO$_4$
scintillators were investigated.
Energy resolution of CdWO$_4$ detector,
coupled to PbWO$_4$ crystal as a light-guide, was
tested. Efficiency of PbWO$_4$-based active shield to suppress
background from the internal contamination of PbWO$_4$ crystals
was calculated. Using of lead tungstate crystal scintillators as high
efficiency $4\pi$ active shield could allow to build sensitive
$2\beta$ experiment with $^{116}$CdWO$_4$ crystal scintillators.

\end{abstract}

\end{frontmatter}

\section{Introduction}

Since 1988 the studies of the $^{116}$Cd $2\beta$ decay have been
performed in the Solotvina Underground Laboratory \cite{SUL} with the
help of cadmium tungstate crystal scintillators (CdWO$_4$) enriched
in $^{116}$Cd to 83\%. The results obtained in the different phases
of these researches have been published earlier \cite{116Cd-K}.
Beginning from 1998, the experiment was carried out in collaboration
with the group from the University and INFN Firenze
\cite{116Cd-KF,W-alpha,Dan03}.

In the apparatus, which is described in detail in
\cite{116Cd-KF,Dan03}, four $^{116}$CdWO$_4$ crystals (total mass 330
g) were exploited. They are viewed by a low background 5'' phototube
(PMT) through light-guide $\oslash 10\times 55$ cm, which is glued of
two parts: quartz (25 cm) and plastic scintillator. The enriched
$^{116}$CdWO$_4$ crystals were surrounded by an active shield made of
15 natural CdWO$_4$ crystals of large volume with total mass of 20.6
kg. These are viewed by a PMT through an active plastic light-guide
$\oslash 17 \times $49 cm. The whole CdWO$_4$ array is situated
within an additional active shield made of plastic scintillator
$40\times 40\times 95$ cm, thus, together with both active
light-guides, a complete $4\pi$ active shield of the $^{116}$CdWO$_4$
detector was provided. Due to the active and passive shields, and as
a result of the time-amplitude and pulse-shape analysis of the data,
the background rate of $^{116}$CdWO$_4$ detectors in the energy
region $2.5-3.2$ MeV ($Q_{2\beta}$ energy of $^{116}$Cd is 2805 keV)
was reduced to 0.04 counts/(yr kg keV). It is one of the lowest
background which has ever been reached with crystal scintillators.
After 14183 h of data taking in the Solotvina Underground Laboratory
the half-life limit on $0\nu2\beta$ decay of $^{116}$Cd was set as
$T_{1/2}\geq1.7\times 10^{23}$ yr at 90\% C.L., which corresponds to
an upper bound on the effective Majorana neutrino mass $\langle m_\nu
\rangle \leq 1.7$ eV \cite{Dan03}. This result is among the strongest
world-wide restrictions (in addition to bounds obtained in
experiments with $^{76}$Ge \cite{Ge76}, $^{82}$Se and $^{100}$Mo
\cite{Arn04}, $^{130}$Te \cite{Te130}, and $^{136}$Xe \cite{Xe136}).
CdWO$_4$ crystals possess several unique properties required for a
$2\beta$ decay experiment: good scintillation characteristics, low
level of intrinsic radioactivity, and possibility of pulse-shape
discrimination to reduce the background.

To enhance sensitivity of $^{116}$Cd experiment to the level of
neutrino mass $0.1-0.05$ eV, one has to increase the measurement time
and the mass of enriched $^{116}$CdWO$_4$, improve the energy
resolution and reduce the background of the detector. As it was
shown by Monte Carlo calculations, the required sensitivity
could be achieved by using ~150 kg of $^{116}$CdWO$_4$ crystals
placed into a large volume of high purity liquid  (CAMEO project
\cite{CAMEO}). The project calls for the background reduction from
the current 0.04 counts/(yr keV kg) to $10^{-3}-10^{-4}$
counts/(yr keV kg). To decrease background, the CAMEO project
intends to use $\approx$1000 t of high purity water or liquid
scintillator ($\approx$$10^{-15}$ g/g for $^{238}$U and $^{232}$Th)
as a shield for $^{116}$CdWO$_4$ crystals. Due to low
density of these liquids, the
necessary dimensions of the shields are huge ($\approx \oslash
12\times 12$ m).

We propose an alternative solution for a sensitive $2\beta$ decay
experiment with $^{116}$CdWO$_4$ by using of lead tungstate
crystal scintillators as high efficiency $4\pi$ active shield.
Lead tungstate (PbWO$_4$) crystal scintillators have been
developed as heavy and fast detectors \cite{PWO-first} for
high-energy physics experiments. Scintillation characteristics of
PbWO$_4$ have been intensively studied during the last decade
\cite{Kob97,Ann00,Kob01,Kob02,Bor04}. High registration efficiency
to $\gamma$ quanta, very good transparency in the region of CdWO$_4$
emission spectrum at the level of a few meters \cite{Bac97},
substantial difference of scintillation decay time
in comparison with CdWO$_4$, well developed tons-scale production
\cite{PWO-production}
make this material very attractive to build a relatively small yet
sensitive experiment to search for $2\beta$ decay of $^{116}$Cd.
In this paper we study the
possibility of applying PbWO$_4$ crystals as material for
light-guides and active shield in a $^{116}$Cd double beta decay
experiment with CdWO$_4$ scintillators.

\section{Measurements and results}

\subsection{Scintillation properties}

The main properties of PbWO$_4$ crystal scintillators are
presented in Table~1 where characteristics of cadmium tungstate
(CdWO$_4$) are also given for comparison. Measurements were carried out
with two clear, colorless, undoped PbWO$_4$ crystals grown by
Czochralski method. One crystal ($45\times 22\times 22$ mm, 182 g
of mass, named PWO-1) was produced in the Institute for
Scintillation Materials (Kharkov, Ukraine), and the second one
($32\times 32\times 10$ mm, 83 g of mass, PWO-2), produced in the
Bogoroditsk Technical Chemical Plant (Russia), was supplied by the
Research Institute of Nuclear Problems (Minsk) \cite{PWO-production}.

\begin{table}[tbp]
\caption{Properties of PbWO$_4$ and CdWO$_4$ crystal
scintillators.}
\begin{center}
\begin{tabular}{|l|l|l|}
\hline
 ~                                     & PbWO$_4$  & CdWO$_4$  \\
\hline
Density (g/cm$^3$)                     &  8.28           & 8.0    \\
Melting point ($^\circ$C)              &  1123           &  1325  \\
Structural type                        &  Sheelite & Wolframite \\
Cleavage plane                         &  Weak (101) & Marked (010) \\
Hardness (Mohs)                        &  3              & $4-4.5$ \\
Wavelength of emission maximum (nm)    & $420-440$       & 480     \\
Refractive index                       &     2.2         & $2.2-2.3$  \\
Effective average decay time$^{\ast}$ ($\mu$s) & 0.01       & 13  \\
Relative photoelectron yield$^{\ast}$  &  6\%     & 100\%    \\
\hline
\multicolumn{3}{l}{$^{\ast}$For $\gamma$ rays, at indoor temperature.} \\
\end{tabular}
\end{center}
\end{table}

Response of the PWO-1 scintillator to $\gamma $ rays was measured
with $^{137}$Cs and $^{207}$Bi $\gamma $ sources at temperatures
$+24^{\circ}$C and $-18^{\circ}$C. The crystal was wrapped by PTFE
reflector tape and optically coupled by Dow Corning Q2-3067
couplant to PMT XP2412. The measured energy spectra are presented
in Fig.~1. The energy resolution FWHM=45\% and 36\% was
obtained at the temperature $-18^{\circ}$C for 570 keV and 662
keV $\gamma$ rays, respectively. The relative pulse amplitude has
been increased in $\approx$3 times with the detector cooling
from $+24^{\circ}$C to $-18^{\circ}$C.

\begin{figure}[htb]
\begin{center}
\mbox{\epsfig{figure=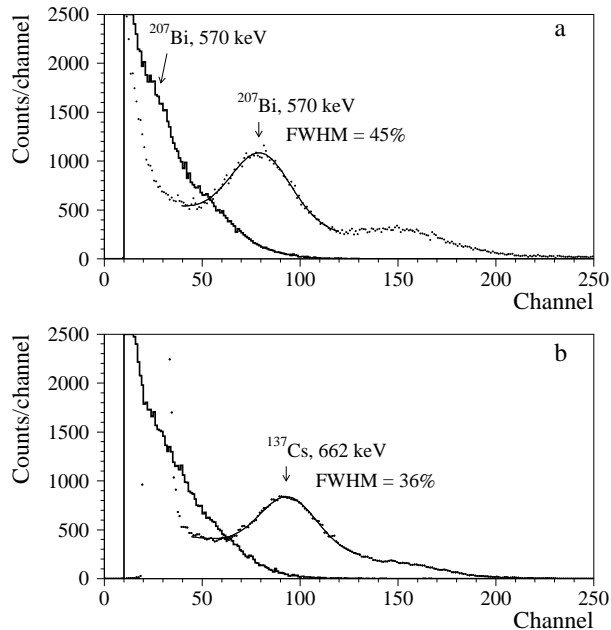,width=8.0cm}}
\caption{Energy spectra measured by PbWO$_4$ scintillation
crystal ($45\times 22\times 22$ mm) with (a) $^{207}$Bi and (b)
$^{137}$Cs $\gamma $ sources at two temperatures: $+24^{\circ}$C
(solid lines) and $-18^{\circ}$C (points).}
\end{center}
\end{figure}

\subsection{$\alpha /\beta $ ratio}

The $\alpha /\beta $ ratio
was measured with the PWO-1 crystal using a collimated $^{241}$Am
$\alpha$ source and thin ($\approx$0.65 mg/cm$^{2}$) mylar
absorbers to obtain $\alpha$ particles in the energy range
$2.1-4.6$ MeV. The energies of $\alpha$ particles were determined
with the help of a surface-barrier detector. In addition, $\alpha$
peak of $^{210}$Po ($E_{\alpha}=5.30$ MeV) from internal
contamination of PbWO$_4$ crystal by $^{210}$Pb (see subsection
2.3) was used. The dependence of the $\alpha/\beta$ ratio on
energy is depicted in Fig. 2 where the $\alpha$ spectrum
measured with the $^{241}$Am alpha source is shown too. The
$\alpha /\beta$ ratio increases above 3 MeV: $\alpha /\beta
=0.08(2)+0.025(5)E_\alpha $, while it decreases as $\alpha /\beta
=0.23(4)-0.024(14)E_\alpha $ at lower energies, where $E_\alpha $
is in MeV. The same behaviour of the $\alpha /\beta $ ratio was
observed for CdWO$_4$ \cite{W-alpha}, calcium tungstate (CaWO$_4$)
\cite{CaWO}, and zinc tungstate (ZnWO$_4$) \cite{ZWO} crystal
scintillators. The energy resolution for $^{210}$Po $\alpha$ peak
was measured as 39\%, which is comparable
with the energy resolution obtained with $\gamma$ sources.

\begin{figure}[htb]
\begin{center}
\mbox{\epsfig{figure=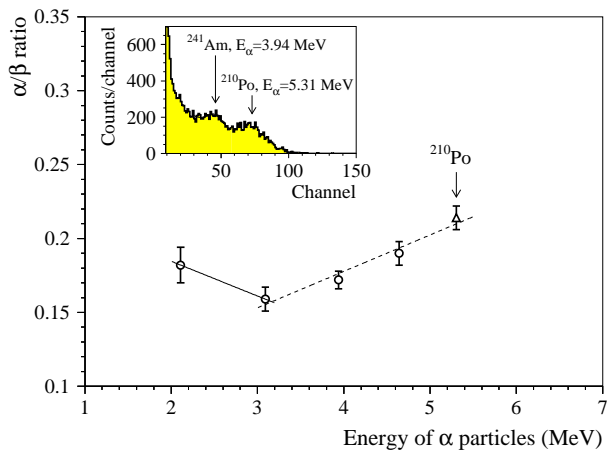,width=8.0cm}}
\caption{Dependence of the $\alpha /\beta $ ratio on energy
measured with the PbWO$_4$ scintillator $45\times 22\times 22$ mm.
The crystal was irradiated by $\alpha $ particles from $^{241}$Am
source through absorbers to obtain energies in $2.1-4.6$ MeV range
(circles). Triangle corresponds to $\alpha$ particles of
$^{210}$Po. (Inset) The $\alpha$ spectrum of $^{241}$Am source
measured with mylar absorber to obtain energy 3.94 MeV. Second
peak at $\approx$73 channel is caused by $\alpha$ decay of
$^{210}$Po (daughter of $^{210}$Pb) inside the scintillator.}
\end{center}
\end{figure}

\subsection{Radioactive contamination of PbWO$_4$ crystal scintillators}

To estimate radioactive contamination, the PWO-1 crystal was
measured in the Solotvina Underground Laboratory built in a salt
mine 430 m underground ($\simeq $1000 m of water equivalent)
\cite{SUL}. The crystal was wrapped by PTFE tape and optically
coupled to low radioactive PMT FEU-110. The detector was cooled to
$-18^{\circ}$C in a temperature-controlled chamber. The shaping
time of the spectroscopy amplifier was set to 0.8 $\mu$s.
Amplitude (energy) and arrival time of signals have been recorded
by the event-by-event data acquisition system. The energy scale
was calibrated with $^{207}$Bi $\gamma$ source. The energy
spectrum accumulated during 2.15 h is shown in Fig. 3. The intense
peak at the energy $\approx$1.2 MeV (in $\gamma$ scale) can be
attributed to intrinsic $^{210}$Po (daughter of $^{210}$Pb from
the $^{238}$U family) with activity of 53(1) Bq/kg. The major part
of events up to the energy $\approx$1 MeV can be ascribed to
$\beta$ active $^{210}$Bi (daughter of $^{210}$Pb). The $^{210}$Pb
contamination of the PWO-2 crystal was measured as 79(3) Bq/kg.

\begin{figure}[htb]
\begin{center}
\mbox{\epsfig{figure=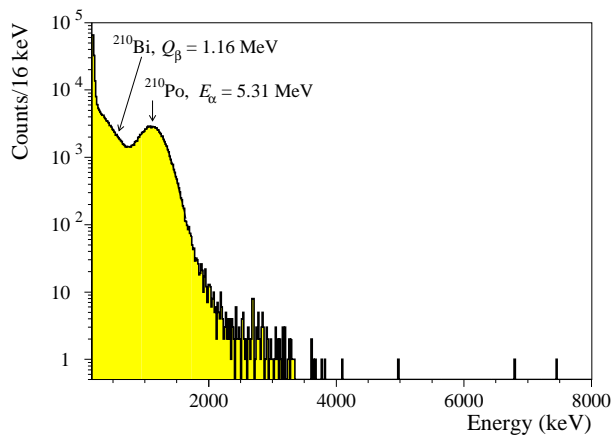,width=8.0cm}}
\caption{Energy spectrum of PbWO$_4$  scintillation crystal
measured in the Solotvina Underground Laboratory at the
temperature $-18^{\circ}$C during 2.15 h. Peak at the energy
$\approx$1.2 MeV (in $\gamma$ scale) can be attributed to
$\alpha$ decay of $^{210}$Po from internal contamination of the
crystal by $^{210}$Pb. Broad distribution up to $\approx$1 MeV
corresponds to $\beta$ spectrum of $^{210}$Bi ($E_{\beta}$=1.16
MeV).}
\end{center}
\end{figure}

Besides, the raw background data accumulated with the PWO-1 were
analyzed by the time-amplitude method (described in detail in
\cite{GSO}), when the energy and arrival time of each event were used
for find the fast sequence of $\beta $ and $\alpha $ decays:
$^{214}$Bi ($Q_\beta =3.27$ MeV) $\rightarrow$ $^{214}$Po ($E_\alpha
=7.69$ MeV, $T_{1/2}=164$ $\mu $s) $\rightarrow$ $^{210}$Pb
($^{238}$U family). To select the $\beta $ decays of $^{214}$Bi,
the energy threshold was set at 0.3 MeV (interval from 0.3 MeV
to the end of the $^{214}$Bi $\beta$ spectrum contains
76\% of the $^{214}$Bi $\beta$ events). For the $\alpha$ decay of
$^{214}$Po, the energy window 1.6--2.6 MeV  (94\% of $\alpha$ events)
and time interval of 90--1000 $\mu $s (67\% of $^{214}$Po decays) were
chosen. There are no peculiarities in the obtained spectra which
could be attributed to the sequence of decays searched for. The limit
on the activity of $^{226}$Ra in the PbWO$_4$ crystal $\leq$10
mBq/kg was set.
Comparing this value with the $^{210}$Po activity, we can conclude that
equilibrium of the uranium chain in
the crystal is strongly broken.

Because of shaping time of the spectroscopy amplifier
(0.8 $\mu$s) exceeds the half-life of $^{212}$Po, and taking into account the $\alpha/\beta$ ratio,
events from the fast sequence of $^{212}$Bi $\beta $ decay ($Q_\beta
=2.25$ MeV) and $^{212}$Po $\alpha $ decay ($E_\alpha =8.78$ MeV,
$T_{1/2}=0.3~\mu$s) can result in one event registered in the detector
with energy from $2.5$ to $5$ MeV. In the energy region 3.4--5 MeV
(where $\approx$60\% of events from the sequence are expected),
there are 7 events, which gives the limit on the activity of
$^{228}$Th ($^{232}$Th family) in the PWO-1 crystal $\leq$13
mBq/kg.

The summary of the measured radioactive contamination of the
PbWO$_4$ crystal scintillators (or limits on their activities) is
given in Table 2 in comparison with the CdWO$_4$ detectors.

\begin{table}[htbp]
\caption{Radioactive contaminations in PbWO$_4$ and CdWO$_4$
crystal scintillators.}
\begin{center}
\begin{tabular}{|l|l|l|l|}
\hline
Chain           & Source        & \multicolumn{2}{|c|}{Activity (mBq/kg)} \\
\cline{3-4}
~               &               & PbWO$_4$             & CdWO$_4$ \cite{Geo96,Bur96,Dan96,Dan03} \\
\hline
~ & ~ & ~ & ~ \\
$^{232}$Th      & $^{228}$Th    & $\leq 13$            & $\leq 0.004-0.039(2)$ \\
~ & ~ & ~ & ~ \\
$^{238}$U       & $^{226}$Ra    & $\leq 10$            & $\leq 0.004$ \\
~               & $^{210}$Pb    & $(53-79)\times 10^3$ & $\leq 0.4$   \\
~ & ~ & ~ & ~ \\
\hline
\end{tabular}
\end{center}
\end{table}

\subsection{PbWO$_4$ crystal as light-guide for CdWO$_4$ scintillator}

A possibility to use PbWO$_4$ crystal as a light-guide for CdWO$_4$
scintillation detector has been tested in measurements. With
this aim, the energy resolution and relative pulse amplitude were
measured with a CdWO$_4$ crystal in two conditions. First, the
CdWO$_4$ crystal ($10\times 10\times 10$ mm, produced in the
Institute for Scintillation Materials, Kharkov), wrapped by PTFE
reflector tape, was optically coupled to PMT XP2412. The
shaping time of the spectroscopy amplifier was set to 16 $\mu$s. The energy
resolution was measured with $^{137}$Cs, $^{207}$Bi, and $^{232}$Th $\gamma $ sources.
In particular, the energy resolution FWHM=3.7\% was obtained for 2615 keV
$\gamma$ line (Fig.~4 (a)). It should be noted that such an energy
resolution was newer reported for CdWO$_4$ scintillation detector.
Then, the CdWO$_4$ crystal was viewed by the PMT through lead
tungstate crystal PWO-1 (wrapped by mylar). The crystals and the
PMT were optically coupled by Dow Corning Q2-3067 couplant.
The energy resolution of 4.1\% (for 2615 keV $\gamma$ line of $^{232}$Th)
and 86\% of relative pulse amplitude were obtained (see Fig.~4 (b)).

\begin{figure}[htb]
\begin{center}
\mbox{\epsfig{figure=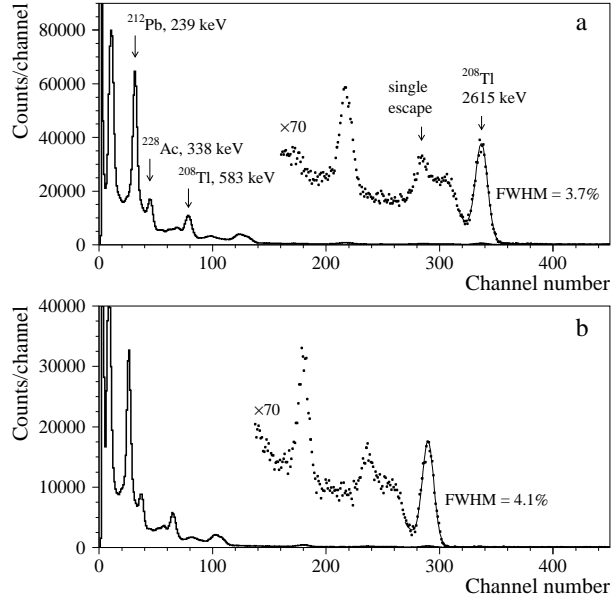,width=8.0cm}}
\caption{Energy spectra measured by CdWO$_4$ scintillation
crystal ($10\times 10\times 10$ mm) with $^{232}$Th $\gamma $
source in two detector arrangements: (a) the crystal wrapped by
PTFE reflector tape optically coupled to PMT; (b) the CdWO$_4$
crystal viewed by the PMT through the PbWO$_4$ crystal
$45\times 22\times 22$ mm as light-guide.}
\end{center}
\end{figure}

\section{Background simulation}

The different sources of background in the $2\beta$ experiment
with $^{116}$CdWO$_4$ crystal scintillators were considered in
\cite{CAMEO}. Here we focus our attention on radioactive
contamination of PbWO$_4$ crystals by $^{232}$Th and $^{238}$U.
Processes with $\beta$, $\alpha$ particles and $\gamma$ rays were
simulated with the help of the GEANT4 package \cite{GEANT} and the
event generator DECAY0 \cite{Pon00}. The
following conditions were taken for the calculations: the CdWO$_4$
crystal ($\oslash$5$\times$5 cm) with the energy resolution (FWHM)
4\% at 2.8 MeV is placed in the center of PbWO$_4$ scintillation
detector ($\oslash$25$\times$25 cm), contaminated by $^{232}$Th
and $^{238}$U at the level of $10^{-12}$ g/g. 7.61$\times$$10^6$
decays of $^{208}$Tl inside the PbWO$_4$ detector
were simulated. It corresponds to exposure of $\approx$220
kg$\times$yr with the CdWO$_4$ detector. The
calculated energy spectrum of the CdWO$_4$ detector, if no
coincidence would be taken into account, is shown in Fig.~5 (a).
The anticoincidence energy spectrum (the energy threshold of
PbWO$_4$ detector was taken to be equal 0.5 MeV) is presented in
Fig.~5 (b). There are only two events in the energy interval of
$0\nu 2\beta$ peak of $^{116}$Cd (2.7--2.9 MeV), which
corresponds to the background counting rate $5\times 10^{-5}$
counts/(yr keV kg) from the $^{232}$Th contamination in PbWO$_4$.

\begin{figure}[htb]
\begin{center}
\mbox{\epsfig{figure=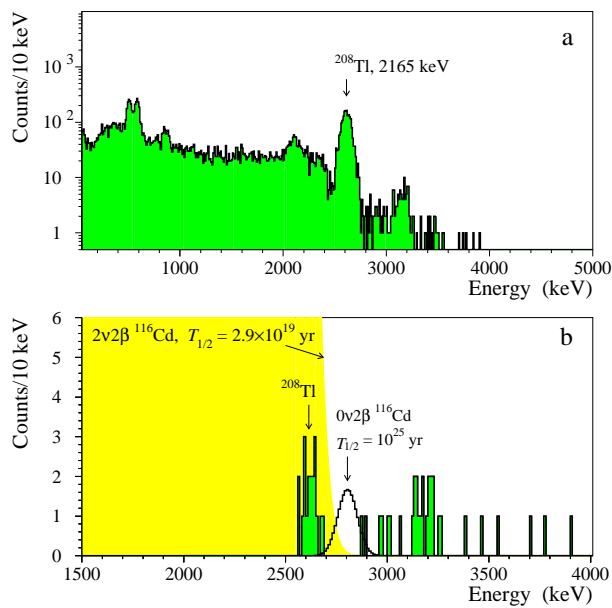,width=8.0cm}}
\caption{Monte Carlo simulated (a) response function of  $^{116}$CdWO$_4$
detector (220 kg$\times$yr of exposure) to decays of $^{208}$Tl
inside shielding PbWO$_4$ crystals (see text). (b) The same as (a)
but in anticoincidence with the PbWO$_4$ detector. Also the $0\nu
2\beta$ peak of $^{116}$Cd with $T_{1/2}=10^{25}$ yr, and two
neutrino $2\beta$ distribution ($T_{1/2}=2.9\times 10^{19}$ yr)
are shown.}
\end{center}
\end{figure}

The contamination of PbWO$_4$ crystals by $^{226}$Ra is even less
dangerous. The Monte Carlo calculations show that no events above
the energy of 2 MeV will be registered in the CdWO$_4$ detector
during $\approx$220 kg$\times$yr of exposure.

\section{Discussion}

The response functions of a detector with enriched
$^{116}$CdWO$_4$ crystals ($\approx$220 kg$\times$yr of exposure) for two
neutrino ($T_{1/2}=2.9\times 10^{19}$ yr \cite{Dan03}) and
neutrinoless $2\beta$ decay of $^{116}$Cd with half-life $10^{25}$ yr are
presented in Fig.~5 (b). Sensitivity of the experiment to neutrinoless
$2\beta$ decay of
$^{116}$Cd is at the level of $\lim T_{1/2}=10^{26}$ yr
(which corresponds to the limit on neutrino mass of $\approx$0.07 eV
\cite{Sta90}). It is evident that the $0\nu2\beta$ decay
with $T_{1/2}\approx 10^{25}$ yr (neutrino mass $\approx$0.2 eV) would be
certainly observed at this level of sensitivity.

The size of PbWO$_4$ active shield\footnote{It should be stressed that CdWO$_4$
crystals, successfully used in the Solotvina experiment \cite{Dan03},
can be also used as active shield detector.} in a set-up could be approximately
of $50\times 50\times 50$ cm. 32 enriched $^{116}$CdWO$_4$ crystals
$\oslash 5\times 5$ cm are viewed by 3'' PMT through logarithmic-spiral
PbWO$_4$ crystals. PbWO$_4$ as light-guide has an advantage
in comparison with plastic or quartz. Because of high index of
refraction (2.2), logarithmic-spiral type light-guide can be made of
less than 60 mm in diameter, that allows to use 3'' PMT (which typically
have lower mass, better energy resolution and lower level of noise)
instead of 5'' PMT in the Solotvina experiment \cite{Dan03}. Assuming
$\approx$50 cm of passive lead shield and $\approx$50 cm polyethylene
antineutron shield, dimensions of the set-up are much more compact
($2.5 \times 2.5 \times 2.6$ m) if to compare with "water shield"
apparatus ($\approx \oslash 12\times 12$ m) proposed in \cite{CAMEO}.

To get such an impressive result, the problem of low-radioactive
PbWO$_4$ crystals production should be solved.
In particular, content of $^{210}$Pb should be decreased: high
counting rate (and, thus, big number of random coincidences) in
$^{210}$Pb--$^{210}$Bi--$^{210}$Po decays, observed in the current
measurements, creates a problem for performing the time-amplitude
analysis of events to search for the specific decay chains.
It is well known that
the freshly smelted lead is contaminated by $^{210}$Pb at the level of
hundreds Bq/kg, while its contamination by uranium and thorium is
substantially less \cite{Bro85}. As the first step, we are going to grow
PbWO$_4$ crystals from archaeological lead aiming to obtain PbWO$_4$
crystals less contaminated by $^{210}$Pb\footnote{For instance,
only limit $\leq$4 mBq/kg on $^{210}$Pb
contamination in lead tungstate crystal produced from Roman lead was
reported in \cite{Ale98}.}. As the next step, we foresee to estimate radioactive
contamination of PbWO$_4$ crystals in low-background measurements by
using the time-amplitude and pulse-shape (to select fast sequence of
$\beta$ and $\alpha$ decays from the  $^{212}$Bi$-^{212}$Po chain)
analyses. As it was demonstrated in the experiments with CdWO$_4$
scintillators, the sensitivities are at the level of a few $\mu$Bq/kg
for $^{228}$Th, $^{226}$Ra, and $^{227}$Ac \cite{Dan03}.

The energy threshold of the shielding PbWO$_4$ detector of 0.5 MeV can
be achieved even with undoped scintillators at room temperature (see
Fig. 1). However, as it was shown in \cite{Ann00}, dopants like
molybdenum and terbium can improve light yield of this scintillator,
which allow to decrease the energy threshold of PbWO$_4$-based active
shield.

\section{Conclusions}

Scintillation properties of PbWO$_4$ crystal scintillators were studied.
The energy resolution FWHM=36\% was measured for the 662 keV $\gamma $ line
of $^{137}$Cs at $-18^{\circ}$C. The $\alpha/\beta$ ratio was measured in the energy
interval 2--5.3 MeV. The dependence of the $\alpha/\beta$ ratio on energy of $\alpha$
particles was observed. Radioactive contamination of two PbWO$_4$ crystals
was measured in the Solotvina Underground Laboratory.
Both crystals are considerably polluted by $^{210}$Po at the level of $50-80$ Bq/kg.
For $^{228}$Th ($^{232}$Th family) and $^{226}$Ra ($^{238}$U) activities only limits
were set at the level of 13 and 10 mBq/kg, respectively.

The excellent energy resolution of FWHM=3.7\% was obtained for 2615 keV
$\gamma$ line with high quality CdWO$_4$ crystal scintillator
$10\times 10\times 10$ mm.
The energy resolution of 4.1\% (2615 keV $\gamma$ line of $^{232}$Th)
and 86\% of relative
pulse amplitude was obtained for CdWO$_4$ scintillator viewed through PbWO$_4$ crystal as
light-guide. We expect an improvement of the light collection and the energy
resolution of
CdWO$_4$ detector by using the logarithmic-spiral PbWO$_4$ light-guide.

Monte Carlo simulation and measurements demonstrate good abilities of PbWO$_4$
crystals to build $4\pi$ active shield for a sensitive $^{116}$Cd double beta
decay experiment with CdWO$_4$ scintillators.

\section{ACKNOWLEDGEMENT}

The authors would like to thank Prof. P.G.~Bizzeti and Prof. P.~Maurenzig from the
Dipartimento di Fisica, Universit\'{a} di Firenze and INFN
(Firenze, Italy) for careful reading of the manuscript and useful comments and
discussion.


\begin{thebibliography}{99}

\bibitem{SUL}  Yu.G.~Zdesenko et al., Proc. 2nd Int. Symp. Underground
                  Physics, Baksan Valley, USSR, August 17--19, 1987 -- Moscow,
                  Nauka, 1988, p. 291.

\bibitem{116Cd-K} F.A.~Danevich et al., Pis'ma Zh. Eksp. Teor. Fiz. 49 (1989) 417
                     [JETP Lett. 49 (1989) 476];\\
                  Yu.G.~Zdesenko, J. Phys. G: Nucl. Part. Phys. 17 (1991) s243;\\
                  F.A.~Danevich et al., Proc. 3-rd Int. Symp. on Weak and
                     Electromagnetic Inter. in Nuclei (WEIN-92), Dubna, Russia,
                     1992 -- World Sci. Publ. Co., 1993, p. 575;\\
                  F.A.~Danevich et al., Phys. Lett. B 344 (1995) 72;\\
                  A.Sh.~Georgadze et al., Phys. At. Nucl. 58 (1995) 1093;\\
                  F.A.~Danevich et al., Nucl. Phys. B (Proc. Suppl.) 48 (1996) 232;\\
                  F.A.~Danevich et al., Nucl. Phys. A 643 (1998) 317;\\
                  F.A.~Danevich et al., Nucl. Phys. B (Proc. Suppl.) 70 (1999) 246.

\bibitem{116Cd-KF} F.A.~Danevich et al., Phys. Rev. C 62 (2000) 045501;\\
                   P.G.~Bizzeti et al., Nucl. Phys. B (Proc. Suppl.) 110 (2002) 389;\\
                   F.A.~Danevich et al., Nucl. Phys. A 717 (2003) 129.

\bibitem{W-alpha} F.A.~Danevich et al., Phys. Rev. C 67 (2003) 014310.

\bibitem{Dan03} F.A.~Danevich et al., Phys. Rev. C 68 (2003) 035501.

\bibitem{Ge76} H.V.~Klapdor-Kleingrothaus et al., Eur. Phys. J. A 12 (2001) 147;\\
               C.E.~Aalseth et al., Phys. Rev. C 59 (1999) 2108;
                  Phys. Rev. D 65 (2002) 092007.

\bibitem{Arn04}  R.~Arnold et al., Pis'ma Zh. Eksp. Teor. Fiz. 80 (2004) 429.

\bibitem{Te130}  C. Arnaboldi et al., Phys. Lett. B 557 (2003) 167; 584 (2004) 260.

\bibitem{Xe136}  R. Luescher et al., Phys. Lett. B 434 (1998) 407;\\
                 R. Bernabei et al., Phys. Lett. B 546 (2002) 23.

\bibitem{CAMEO} G. Bellini et al., Phys. Lett. B 493 (2000) 216;
                   Eur.~Phys.~J. C 19 (2001) 43.

\bibitem{PWO-first} V.G.~Barishevsky et al., Nucl. Instr. and
                       Meth. A 322 (1992) 231.

\bibitem{Kob97} M. Kobayashi et al., Nucl. Instr. Meth. A 399 (1997) 261.

\bibitem{Ann00} A. Annenkov et al., Nucl. Instr. Meth. A 450 (2000) 71.

\bibitem{Kob01} M. Kobayashi et al., Nucl. Instr. Meth. A 465 (2001) 428.

\bibitem{Kob02} M. Kobayashi et al., Nucl. Instr. Meth. A 484 (2002) 140.

\bibitem{Bor04} A.~Borisevich et al., Nucl. Instr. Meth. A, in press.

\bibitem{Bac97} S.~Baccaro et al., Nucl. Instrum. Meth. A 385 (1997) 209.

\bibitem{PWO-production} A. Annenkov et al., Nucl. Instr. Meth. A, in press.

\bibitem{CaWO} Yu.G.~Zdesenko et al., Nucl. Instr. Meth. A, in press.

\bibitem{ZWO} F.A.~Danevich et al., nucl-ex/0409014, submitted to Nucl. Instr. Meth.  A.

\bibitem{GSO}  F.A.~Danevich et al., Nucl. Phys. A 694 (2001) 375.

\bibitem{Geo96}  A.Sh.~Georgadze et al., Instr. Exp. Technique 39 (1996) 191.

\bibitem{Bur96} S.Ph.~Burachas et al., Nucl. Instr. Meth. A 369 (1996) 164.

\bibitem{Dan96}  F.A.~Danevich et al., Z. Phys. A 355 (1996) 433.

\bibitem{GEANT} S.~Agostinelli et al. (GEANT4 Collaboration).
                   Nucl. Instr. Meth. A 506 (2003) 250;
                   http://geant4.web.cern.ch/geant4/.

\bibitem{Pon00} O.A.~Ponkratenko et al., Yad. Fiz. 63 (2000) 1355
                   [Phys. Atom. Nucl. 63 (2000) 1282].

\bibitem{Sta90} A.~Staudt et al., Europhys. Lett. 13 (1990) 31.

\bibitem{Bro85} R.L.~Brodzinski et al., Nucl. Instr. Meth. A 239 (1985) 207.

\bibitem{Ale98} A.~Alessandrello et al., Nucl. Instr. Meth. A 409 (1998) 451.

\end{thebibliography}
\end{document}